\documentclass{emulateapj}

\usepackage{graphics}
\usepackage{epsfig}
\usepackage{natbib}
\shorttitle{Far-UV Continuum in Disks II: CO}
\shortauthors{France et al.}

\begin{document}

\title{The Far-Ultraviolet ``Continuum'' in Protoplanetary Disk Systems II: 
CO Fourth Positive Emission and Absorption\altaffilmark{*}}

\author{
Kevin France\altaffilmark{1}, Eric Schindhelm\altaffilmark{1}, Eric B. Burgh\altaffilmark{1},
Gregory J. Herczeg\altaffilmark{2}, Graham M. Harper\altaffilmark{3}, Alexander Brown\altaffilmark{1}, James C. Green\altaffilmark{1}, Jeffrey L. Linsky\altaffilmark{4}, 
Hao Yang\altaffilmark{4}
Herv{\'e} Abgrall\altaffilmark{5},
David R. Ardila\altaffilmark{6},
Edwin Bergin\altaffilmark{7},  Thomas Bethell\altaffilmark{7}, Joanna M. Brown\altaffilmark{2},
Nuria Calvet\altaffilmark{7}, 
Catherine Espaillat\altaffilmark{8},
Scott G. Gregory\altaffilmark{9}, Lynne A. Hillenbrand\altaffilmark{9}, Gaitee Hussain\altaffilmark{10},
Laura Ingleby\altaffilmark{7}, Christopher M. Johns-Krull\altaffilmark{11}, Evelyne Roueff\altaffilmark{5}, Jeff A. Valenti\altaffilmark{12}, Frederick M. Walter\altaffilmark{13}
}


\altaffiltext{*}{Based on observations made with the NASA/ESA $Hubble$~$Space$~$Telescope$, obtained from the data archive at the Space Telescope Science Institute. STScI is operated by the Association of Universities for Research in Astronomy, Inc. under NASA contract NAS 5-26555.}


\begin{abstract}
We exploit the high sensitivity and moderate spectral resolution of the $HST$-Cosmic Origins Spectrograph to detect far-ultraviolet spectral features of carbon monoxide (CO) present in the inner regions of protoplanetary disks for the first time. We present spectra of the classical T Tauri stars HN Tau, RECX-11, and V4046 Sgr, representative of a range of CO radiative processes. 
HN Tau shows CO bands in absorption against the accretion continuum.  The CO absorption  
most likely arises in warm inner disk gas.  We measure a CO column density and rotational excitation temperature of $N$(CO) = 2~$\pm$~1~$\times$~10$^{17}$ cm$^{-2}$ and $T_{rot}$(CO) 500~$\pm$~200~K for the absorbing gas. 
We also detect CO $A$~--~$X$ band emission in RECX-11 and V4046 Sgr, excited by ultraviolet line photons, predominantly \ion{H}{1} Ly$\alpha$.  All three objects show emission from CO bands at $\lambda$~$>$ 1560~\AA, which may be excited by a combination of UV photons and collisions with non-thermal electrons.  In previous observations these emission processes were not accounted for due to blending with emission from the accretion shock, collisionally excited H$_{2}$, and photo-excited H$_{2}$; all of which appeared  as a ``continuum'' whose components could not be separated.  The CO emission spectrum is strongly dependent upon the shape of the incident stellar Ly$\alpha$ emission profile. We find CO parameters in the range: $N$(CO)~$\sim$~10$^{18-19}$ cm$^{-2}$, $T_{rot}$(CO)~$\gtrsim$~300~K for the Ly$\alpha$-pumped emission. 
We combine these results with recent work on photo- and collisionally-excited H$_{2}$ emission, concluding that the observations of ultraviolet-emitting CO and H$_{2}$ are consistent with a common spatial origin. 
We suggest that the CO/H$_{2}$ ratio ($\equiv$~$N$(CO)/$N$(H$_{2}$)) in the inner disk is $\sim$~1, a transition between the much lower interstellar value and the higher value observed in solar system comets today, a result that will require future observational and theoretical study to confirm.  
\end{abstract}


\section{Introduction}


Observations of classical T Tauri star (CTTS) disks offer a snapshot of the formation epoch of Jovian extrasolar planets. While the majority of the mass in these young disks (age~$\lesssim$~10 Myr) is in the form of molecular hydrogen (H$_{2}$), the lack of a permanent dipole moment makes H$_{2}$ gas difficult to observe in the cool outer regions of the protoplanetary disk. Millimeter (mm) and sub-mm observations of CO can be used to trace molecular material in the outer disk
(e.g., Dutrey et al. 1996; Qi et al. 2004), while near- and mid-infrared (IR) observations of OH, CO, CO$_{2}$, H$_{2}$O, and other molecules are powerful diagnostics of inner gas disks ($a$~$<$~5AU), the region of terrestrial planet formation~\citep{najita03,carr04,salyk07,salyk08,carr08,bethell09}.~\nocite{dutrey96,qi04} 
IR observations of the overtone ($\Delta$$v$~=~2) and fundamental ($\Delta$$v$~=~1) bands of CO have been the most widely used tracers of the inner gas disk. 
The overtone bands are thought to trace the atmosphere of the optically thick gas 
at the very inner regions of the disk ($a$~$\lesssim$~0.2 AU), where the 
temperature and densities are high ($T$~$\approx$~2000~--~3000 K, 
$n_H$~$>$~10$^{10}$ cm$^{-3}$; Najita et al. 1996; Carr \& Najita 2004).~\nocite{najita96,carr04} 
The fundamental CO emission spectrum traces gas at $T$~$\approx$ 1000~--~2000 K and radii 0.04~$<$~$a$~$\lesssim$~1 AU~\citep{najita03,najita07}, possibly arising in a disk atmosphere or regions of lower column density cleared out by dynamical processes (e.g., binary stellar systems or planetary interactions). 

Far-ultraviolet (UV) observations of CTTSs reveal a wealth of emission lines arising from H$_{2}$, whose electronic transition spectrum is dipole-allowed. Far-UV H$_{2}$ was first identified in CTTSs by~\citet{brown81}, and subsequent observations at higher spectral resolution~\citep{herczeg02,ardila02,herczeg06} have shown that this emission arises in a photo-excited (``pumped'') warm surface layer of the disk~\citep{herczeg04}, or in extended outflows~\citep{walter03,saucedo03}.   
Recent observations have confirmed this picture, detecting both the fluorescent emission lines of H$_{2}$ and the absorption of pumping photons from the disk-reflected Ly$\alpha$ profile~\citep{yang11}. Collisional excitation of H$_{2}$ by non-thermal electrons has been proposed to explain the faint 1300~--~1650~\AA\ emission in CTTS disks~\citep{bergin04}. Subsequent work has shown this excess to be ubiquitous~\citep{ingleby09}, but the low spectral resolution and high instrumental backgrounds of the {\it Hubble Space Telescope}-ACS and -STIS, respectively, could not cleanly separate the components of this far-UV ``continuum''. 
France et al. (2011), hereafter Paper I, used the dramatic gains in spectroscopic sensitivity offered by $HST$-COS to separate and identify three components of the faint excess emission. We use the term ``continuum'' because these emissions were mostly unresolved in previous studies. The purpose of this series of papers is to distinguish continuum and line-emission processes in CTTS spectra, and use this information to better understand two regions where far-UV photons are emitted: the accretion shock and the inner molecular disk. 
Paper I describes the accretion continuum and electron-excited H$_{2}$ spectrum for two prototypical objects. In the present work, we present the first detections of emission and absorption from the CO Fourth Positive band system ($A$$^{1}\Pi$~--~$X$$^{1}\Sigma^{+}$)  in protoplanetary disks. These bands are observed in the wavelength range 1270~$\lesssim$~$\lambda$~$\lesssim$~1720~\AA.

While emission and absorption from the Fourth Positive band system are widely-used molecular tracers in other areas of astrophysics, this is the first instance to our knowledge where these lines have been seen in young, low-mass protoplanetary disks. We provide this very brief review of $A$~--~$X$ observations in other astrophysical environments because the literature in these fields is considerably more developed than what exists for protoplanetary disks.  
The absorption bands of CO have been long studied in the ISM~\citep{federman80,morton94,burgh07,sheffer08}. Absorption lines from cold CO have been seen in older debris disks (e.g., $\beta$ Pic, age~$\sim$~8~--~20~Myr; Vidal-Madjar et al. 1994; Roberge et al. 2000) and in the disk of the Herbig Ae star AB Aur (age~$\approx$~2 Myr; Roberge et al. 2001). In these cases, the CO is too cold to be in the inner disk, and may be replenished by the collision of planetesimals. The $A$~--~$X$ emission bands of CO are prominent in the far-UV spectra of comets~\citep{feldman76,mcphate99,lupu07}, Venus~\citep{durrance81,hubert10}, and Mars~\citep{feldman00,krasnopolsky02}. Emission and absorption spectra of CO have been observed in the atmospheres of cool stars~\citep{carpenter94,hinkle05}, including the Sun~\citep{goldberg65,bartoe78}, and dominate the far-UV spectrum of the Red Rectangle (HD44179; see Sitko et al. 2008 and references therein).\nocite{sitko08} In this work we describe how the far-UV emission and absorption bands of CO can be used to constrain the molecular properties of protoplanetary disks.~\nocite{vidal94,roberge00,roberge01} 

We use three objects as prototypical examples of the range of CO spectral signatures in these objects.  
 The spectra can be categorized as CO emission dominated by strong Ly$\alpha$ pumping (V4046 Sgr), CO emission coming from a combination of Ly$\alpha$ and \ion{C}{4} pumping photons with a possible contribution from electron-impact excitation (RECX-11), and CO absorption through a disk (HN Tau). In \S2, we describe the targets and COS observations. In \S3, we describe the qualitative features of the CO spectrum and present simple models for the emission and absorption. We use these results to constrain the physical conditions of the inner molecular disk and the CO/H$_{2}$ ratio in \S4. We present a brief summary of the paper in \S5. 

\section{Targets and Observations} 
HN Tau, RECX-11, and V4046 Sgr are pre-main-sequence systems with actively accreting disks. 
These targets are found in associations typical of ``young'' (HN Tau) and ``old'' (RECX-11 and V4046 Sgr) disk populations.  
HN Tau is a 3.1\arcsec\ ($\approx$~430 AU) separation binary consisting of a K5 primary and mid-M secondary with masses of~0.8 and 0.2~$M_{\odot}$~\citep{white01}. 
HN Tau A was the only stellar component in the spectroscopic aperture used in this work.  This system is a member of the well-studied Taurus star-forming complex at a distance of $d$~$\sim$~140 pc. HN Tau has strong outflows that have been studied extensively with optical spectroscopy~\citep{hartigan04}. 
The typical age for young, gas- and dust-rich disks in the Taurus region is $\sim$~1~--~2 Myr~\citep{kenyon95}.  
This is roughly consistent with the work of~\citet{kraus09}, who find an ages of $\approx$~2 and 4 Myr for HN Tau A and B, respectively.  
HN Tau is heavily veiled in the optical, and observations of the outer dust disk at 850~$\mu$m indicate a dust mass of ~$\approx$~270~$M_{\oplus}$~\citep{andrews05}.  
The inclination of HN Tau is not known.

RECX-11 is a K5.5 candidate  CTTS~\citep{luhman04} with a mass of 0.9~$M_{\odot}$~\citep{mamajek99}.  The disk inclination is estimated to be $i$~$\approx$~70\arcdeg\ based on magnetospheric accretion model  fits to the H$\alpha$ line profile observed by~\citet{lawson04}.  It is a member of the $\eta$ Cha star cluster, located at a distance of 97 pc~\citep{mamajek99}. 
The cluster is estimated to have an age of 5~--~8~Myr~\citep{mamajek99,luhman04}, suggesting that the disk systems in $\eta$ Cha are likely to be in the process of dissipating their primordial gas envelopes, however RECX-11 displays a near-IR excess that is consistent with gas-rich, dusty CTTS disks in the younger Taurus cluster~\citep{aguilar09}. The H$\alpha$ profile of RECX-11 suggests that it could be an accreting system~\citep{rayjay06}, and as we will show in \S3.2, this object displays far-UV continuum emission characteristic of weak accretion. Similarly,~\citet{ingleby11} use a combination of far- and near-UV spectra to infer active accretion onto RECX-11, confirming its status as a CTTS. 
RECX-11 was part of a near-IR survey to measure the H$_{2}$ content of the inner disk, but no warm ($T$~$\sim$~2000~K) H$_{2}$ was detected~\citep{howat07}. The non-detection of the near-IR rovibrational emission of H$_{2}$, contrasted with the $\sim$~100 photo-excited H$_{2}$ emission lines observed in the COS data (Ingleby et al. 2011; this work) is a powerful demonstration of the utility of far-UV molecular observations in the study of the inner regions of protoplanetary disks. 

\begin{deluxetable*}{ccccccc}
\tabletypesize{\scriptsize}
\tablecaption{$HST$-COS observing log. \label{cos_obs}}
\tablewidth{0pt}
\tablehead{
\colhead{Object} & \colhead{R. A. (J2000)} & \colhead{Dec. (J2000) }& \colhead{Date} & \colhead{COS Modes} & \colhead{T$_{exp}$[G130M] (s)} & \colhead{T$_{exp}$[G160M] (s)} 
}
\startdata 
RECX-11 & 08$^{\mathrm h}$ 47$^{\mathrm m}$ 01.28$^{\mathrm s}$ & -78\arcdeg\ 59\arcmin\ 34.1\arcsec & 2009 Dec 12 & G130M, G160M & 3645 & 4514 \\

HN Tau & 04$^{\mathrm h}$ 33$^{\mathrm m}$ 39.37$^{\mathrm s}$ & +17\arcdeg\ 51\arcmin\ 52.1\arcsec & 2010 Feb 10 & G130M, G160M & 5725 & 4529  \\

V4046 Sgr & 18$^{\mathrm h}$ 14$^{\mathrm m}$ 10.49$^{\mathrm s}$ & -32\arcdeg\ 47\arcmin\ 34.2\arcsec & 2010 Apr 27 & G130M, G160M & 4504  & 5581 
\enddata
\end{deluxetable*}

\begin{figure}[h]
\begin{center}
\epsfig{figure=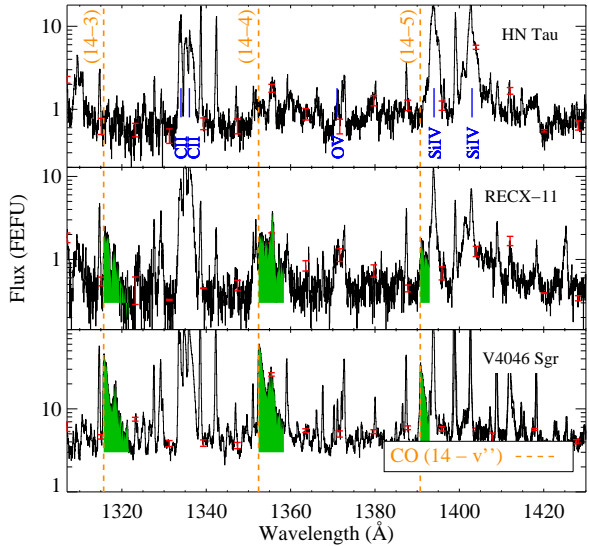,width=2.55in,angle=90}
\vspace{+0.2in}\caption{\label{cosovly} The 1307~--~1430~\AA\ spectra of HN Tau, RECX-11, and V4046 Sgr; roughly in order of increasing age from top to bottom.  All three spectra show an underlying continuum and emission from atomic species, labeled in blue.  The spectrum of V4046 Sgr shows strong CO emission from the (14~--~$v^{''}$) bands in this wavelength interval.   We attribute this emission to pumping by stellar Ly$\alpha$ photons.  The (14~--~$v^{''}$) bandheads for $v^{''}$ = 3, 4, and 5 are marked with orange dashed lines. 
Emission lines not labeled as CO or atomic in this wavelength region are produced by photo-excited H$_{2}$~\citep{herczeg02,yang11}. 
 RECX-11 shows less Ly$\alpha$-pumped CO than V4046 Sgr and we do not detect this emission toward HN Tau.  The flux is shown on a log scale so the emission from each spectral component can be displayed on a single figure.  
The flux is plotted in FEFU ($\equiv$~femto-erg flux unit, 1 FEFU~=~1 $\times$~10$^{-15}$ erg cm$^{-2}$ s$^{-1}$ \AA$^{-1}$).  Representative error bars on the flux are shown in red.}
\end{center}
\end{figure}

V4046 Sgr is a close binary ($r$~$\approx$~9~$R_{\odot}$) composed of two mid-K stars (0.91 and 0.87~$M_{\odot}$; Stempels \& Gahm 2004) with a circumbinary disk at an intermediate inclination angle (binary system and disk inclinations of $i$~$\sim$~35\arcdeg, Quast et al. 2000; Rodriguez et al. 2010).\nocite{quast00,rodriguez10} It is one of the brightest CTTSs in the UV and X-ray bands~\citep{gunther06,gunther08}, located at a distance of 72 pc~\citep{torres06}.  
It has an age between 4~--~12 Myr, depending upon whether or not it is a member of the $\beta$~Pic moving group~\citep{quast00,kastner08}. V4046 Sgr has a molecule-rich outer disk ($M_{gas}$~$\sim$~110~$M_{\oplus}$, $M_{dust}$~$\sim$~40~$M_{\oplus}$; Rodriguez et al. 2010), however  $\lambda$~$<$~10~$\mu$m photometry strongly suggests that the inner disk has been cleared of dust~\citep{jensen97}. 
V4046 Sgr was studied as part of the $IUE$ pre-main sequence star atlas~\citep{krull00}, and while H$_{2}$ emission from the Lyman and Werner levels is strong (see e.g., Paper I), a mid-IR search for emission from cooler H$_{2}$ has returned only upper limits~\citep{carmona08}.
The accretion continuum and electron impact excited H$_{2}$ emission from V4046 Sgr were presented in Paper I. \nocite{france11a}

Our targets were observed with the medium-resolution, far-UV (G130M and G160M) modes of COS~\citep{osterman11} during cycle 17, with HN Tau and RECX-11 observed under $HST$ program 11616 (P.I.~--~G. Herczeg) and V4046 Sgr observed as part of the COS GTO program (P.I.~--~J. Green). 
The total observing times for the three objects were 10.3 ks, 8.2 ks, and 10.1 ks, respectively. 
In order to achieve continuous spectral coverage and minimize fixed pattern noise,
observations in each grating were made with several central wavelength and focal-plane positions (FP-POS). 
This combination of grating settings covers the 1140~$\leq$~$\lambda$~$\leq$~1760~\AA\ bandpass for all targets, at a resolving power of $R$~$\approx$~18,000\footnote{The COS LSF experiences a wavelength dependent non-Gaussianity due to the introduction of mid-frequency wave-front errors produced by the polishing errors on the $HST$ primary and secondary mirrors; {\tt http://www.stsci.edu/hst/cos/documents/isrs/}. We note that for broad emission lines ($FWHM$~$\gtrsim$~75 km s$^{-1}$), the linespread function is essentially indistinguishable from Gaussian~\citep{france10b}.}. Near-UV imaging target acquisitions were performed through the COS primary science aperture using MIRRORA for HN Tau and MIRRORB for RECX-11 and V4046 Sgr. 
Table 1 provides a log of the COS observations acquired as part of this study. The data have been processed with the COS calibration pipeline, CALCOS\footnote{We refer the reader to the COS Instrument Handbook for more details: {\tt http://www.stsci.edu/hst/cos/documents/handbooks/current/cos\_cover.html}}, and combined with the custom IDL coaddition procedure described by~\citet{danforth10} and~\citet{shull10}. 

\begin{figure}
\begin{center}
\epsfig{figure=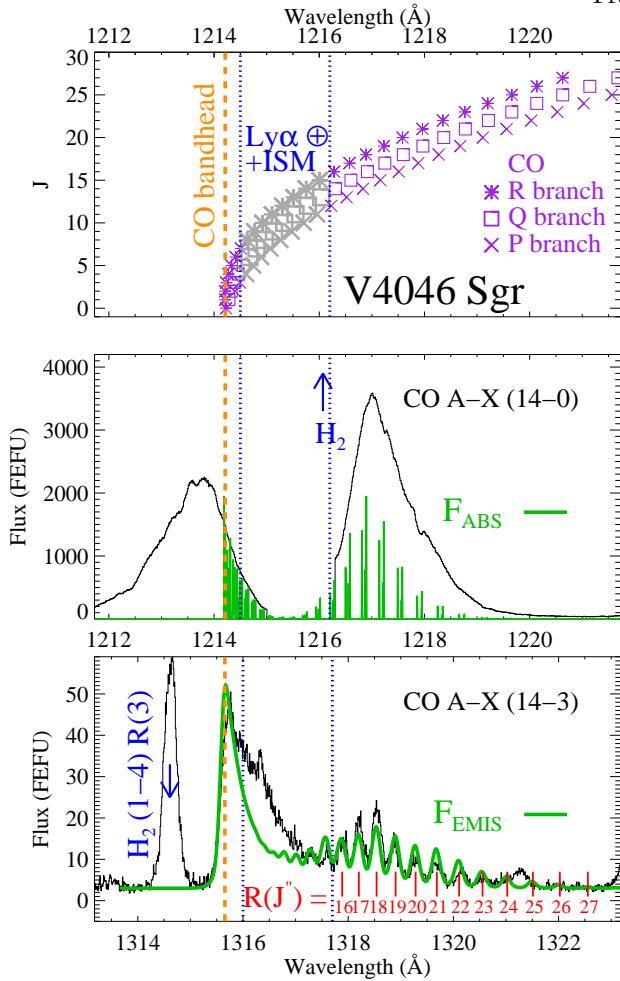,width=4.75in,angle=90}
\vspace{+0.1in}
\caption{
\label{cosovly} A graphical description of the \ion{H}{1} Ly$\alpha$ pumping line in V4046 Sgr.
The top plot shows the distribution of rotational states (marked with purple stars, squares, and xs) of the CO $A$~--~$X$ (14~--~0) band.  The rotational states ($J$) corresponding to the center of the stellar+shock Ly$\alpha$ profile are marked by the larger gray symbols; the exciting radiation field at these wavelengths is highly uncertain due to outflow+interstellar absorption and geocoronal emission (labeled ``Ly$\alpha$~$\oplus$ + ISM'').  We do not attempt to reconstruct the Ly$\alpha$ line profile at these wavelengths, instead we fit a parabola through this region in order create a continuous pumping profile.  In the middle and bottom panels, we show the observed V4046 Sgr Ly$\alpha$ emission line and the (14~--~3) CO emission band.  
The (14~--~0) and (14~--~3) $A$~--~$X$ bandheads are marked by the dashed orange lines.
The green curves show our model (\S3.3.1) for the total Ly$\alpha$ flux absorbed by CO ($middle$) and the fluorescent emission in the (14~--~3) band ($bottom$). 
We find a best-fit model with $N$(CO)~=~5.6~$\pm$~2.3~$\times$~10$^{18}$ cm$^{-2}$ and 
$T_{rot}$(CO) = 290~$\pm$~90 K.    The high rotational level R branch lines are labeled in red on the bottom plot, showing the extent of the observed rotational distribution.
 The bottom panel shows a strong photo-excited line of H$_{2}$, $B$~--~$X$ (1~--~4) R(3).  This line is also pumped by Ly$\alpha$, and we mark the pumping  transition wavelength with an ``up'' arrow in the middle panel. 
 }
\end{center}
\end{figure}

\section{Results and Analysis}

\subsection{Observations of the Carbon Monoxide Fourth Positive System }

The three systems discussed here represent the range of CO signatures observed in CTTS spectra during the first year of COS observations.
A complete survey of the CO emission observed in the combined GO and GTO observing programs, including spectral modeling to determine CO column densities, CO rotational temperatures, and the spatial distribution of far-UV CO emission for approximately a dozen CTTS disks, is in preparation. Figure 1 shows the spectra in the 1307~--~1430~\AA\ bandpass. The spectra are displayed on a log scale so that the emission from CO, photo-excited H$_{2}$, and atomic species can be shown on a single plot.  The flux units used on this and subsequent figures are FEFUs\footnote{the Femto-Erg Flux Unit; see \S1.1.2 of the Cycle 17 COS Instrument Handbook} (1 FEFU~=~1 $\times$~10$^{-15}$ ergs cm$^{-2}$ s$^{-1}$ \AA$^{-1}$) . 
All three objects show the accretion-generated continuum described in Paper I (see also Calvet \& Gullbring 1998 for a detailed theoretically-based discussion of the ultraviolet-optical spectrum of the accretion shock). 
The atomic emission lines formed in the active atmosphere~\citep{bouvier07}, funnel flows (e.g. Muzerole et al. 2001), and the accretion shock~\citep{gunther08} are labeled in blue.\nocite{calvet98} In this wavelength region, \ion{C}{2} $\lambda$1334, 1335~\AA\ and \ion{Si}{4} $\lambda$1394, 1403~\AA\ are the strongest atomic species. In the following subsections, we describe the major observational result of this work: the first detection of prominent far-UV emission and absorption features of CO in the spectra of CTTSs.~\nocite{muzerole01}

\subsubsection{Ly$\alpha$ Pumping of CO: Selective Rotational Photo-Excitation}

After the atomic emission lines mentioned above, the next strongest broad emission lines observed are those of CO.  
The bright, wide, sawtooth emission features seen in the spectrum of V4046 Sgr ({\it Figure 1; bottom panel} and {\it Figure 2; bottom panel}) are produced by the $A$~--~$X$ (14~--~$v^{''}$) transitions of CO\footnote{The quantum numbers $v$ and $J$ denote the ground electronic state ($X$$^{1}\Sigma^{+}$), the numbers $v^{'}$ and $J^{'}$ characterize the CO in the excited ($A$$^{1}\Pi$) electronic state, and the numbers $v^{''}$ and $J^{''}$ are the rovibrational levels of the electronic ground state  following the fluorescent emission.  Absorption lines are described by ($v^{'}$~--~$v$) and emission lines by ($v^{'}$~--~$v^{''}$).}.
The bandheads for the $v^{''}$~=~3, 4, and 5 transitions (1315.7, 1352.4, and 1390.7~\AA) are marked.
The rotational states of each vibrational band pile up at the bandhead and extend to the red with increasing rotational level.  

This (14~--~$v^{''}$) progression is pumped through a wavelength coincidence with the stellar Ly$\alpha$ profile. The bandhead of the (14~--~0) band is at $\lambda$~=~1214.2~\AA, with higher rotational states extending redward across the \ion{H}{1} emission line. This Ly$\alpha$ pumped emission is analogous to the fluorescent H$_{2}$ cascades pumped by Ly$\alpha$ seen in CTTS spectra~\citep{herczeg06}. Many emission lines from the Ly$\alpha$ pumped H$_{2}$ cascade are also visible in the spectra displayed in Figure 1. The CO emission spectrum is present, but less strong in RECX-11, and is not detected in HN Tau. The width of each observed band is 6~--~7~\AA, indicating that rotational excitation as high as $J$~$\approx$~25 is present.  It is possible that higher rotational states are populated but not observed in emission because of a lack of Ly$\alpha$ pumping flux redward of 1218~\AA.  

We detect 8 bands of the (14~--~$v^{''}$) progression in V4046 Sgr: all bands with R(0)  Einstein $A$-values~$\gtrsim$~2~$\times$~10$^{6}$ s$^{-1}$ from (14~--~2) $\lambda$1280~\AA\ to (14~--~12) $\lambda$1713~\AA.  We detect the complete band sequence with line ratios consistent with the expected branching ratios, demonstrating the solidity of the spectroscopic identifications.  
We observe (14~--~$v^{''}$) bands with $v^{''}$ = 2, 3, 4, 5, 7, and 12 in the spectrum of RECX-11.  Bands with $v^{''}$ = 1 and 8~--~11 are not detected due to small branching ratios and spectral overlap with photo-excited H$_{2}$.  
The $v^{''}$~=~3, 4, and 5 bands displayed in 
Figure 1 are the most readily observable features as these have the largest branching ratios and lie nearest to the peak of the COS spectroscopic sensitivity~\citep{osterman11}.  
Table 2 gives a summary of the detected CO emission features. 

\begin{figure*}
\begin{center}
\epsfig{figure=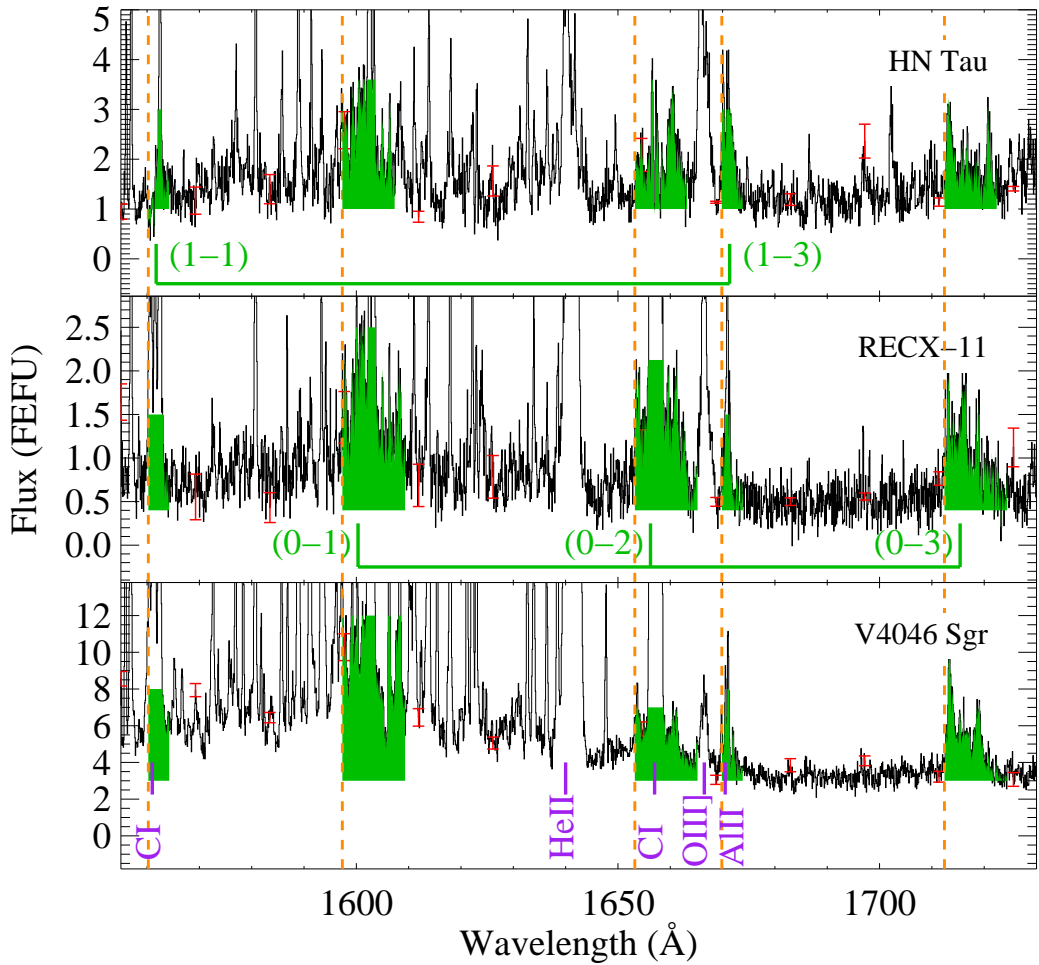,width=4.5in,angle=90}
\vspace{+0.5in}
\caption{ A montage of the CO $A$~--~$X$ emission in the 1560~--~1730~\AA\ wavelength interval.  
The flux attributable to CO is shaded in green.  
The emission at 1713~\AA\ is a combination of the (14~--~12) and (0~--~3) progressions in V4046 Sgr and RECX-11.  The 1713~\AA\ lines are an important signpost for the presence of CO in the spectrum because this wavelength region is free of blending with H$_{2}$. 
CO bandheads are marked with dashed orange lines, and atomic features are labeled in purple.  CO line identifications and possible excitation mechanisms are presented in Table 2.  All other emission features not identified as CO or atomic emission in this figure are
photo- and collisionally-excited H$_{2}$ (Herczeg et al. 2002; Paper I).  
\label{cosovly} 
 }
\end{center}
\end{figure*}

Figure 2 ($bottom$) displays the (14~--~3) band of CO observed in V4046 Sgr in greater detail, as well as the stellar Ly$\alpha$ profile (Figure 2, $middle$) obtained in the same observations. 
The CO emission displays rotational structure that reflects the temperature of the CO-bearing gas and the shape of the exciting Ly$\alpha$ radiation field. The intermediate rotational levels are seen to be suppressed (but not absent) in the CO emission spectrum, indicating a self-reversal of the Ly$\alpha$ line profile. This would be expected if the core of the Ly$\alpha$ experiences self-absorption either from an outflow or from neutral hydrogen in the upper disk layers. 
This is qualitatively similar to the inferred self-reversal of the Ly$\alpha$ profile at the warm H$_{2}$ layer of the TW Hya disk~\citep{herczeg04}.   
We cannot directly measure the core of the Ly$\alpha$ emission line due to absorption from the intervening ISM and outflows, as well as contamination from geocoronal Ly$\alpha$ that fills the COS aperture (see \S3.2). The region over which geocoronal and interstellar Ly$\alpha$ contaminate the V4046 Sgr spectrum is marked with dotted lines in Figure 2 ($middle$). 
We calculate the total Ly$\alpha$-pumped CO emission ($F(CO)$, in units of erg cm$^{-2}$ s$^{-1}$) by integrating the 
(14~--~3) band ($F(14-3)$) over the 6~\AA\ redward of the bandhead, subtracting the 
continuum contribution over the same region.  The total CO emission is 
$F(CO)$~=~$F(14-3)$/$B(14-3)$, where $B(14-3)$ is the branching ratio for 
emission into the $v^{''}$~=~3 level.  The total Ly$\alpha$-pumped CO fluxes
are listed in Table 3. 

Figure 2 also shows a complementary example of Ly$\alpha$ pumped H$_{2}$ emission from the V4046 Sgr disk, the $B$~--~$X$ (1~--~4) R(3) [$\lambda_{rest}$~=~1314.62~\AA] line, which is pumped by the coincidence of Ly$\alpha$ with the (1~--~2) P(5) [$\lambda_{rest}$~=~1216.07~\AA] transition. The absorbing transition is marked with an ``up'' arrow, and the emission line is marked with a ``down'' arrow. 

\subsubsection{Other CO Emission Bands}

We also identify CO $A$~--~$X$ emission excited through other routes, most notably the (0~--~1) $\lambda$1597~\AA\ and (0~--~3) $\lambda$1712~\AA\ bands in all three systems (Figure 3). These emissions may be attributed to pumping by \ion{C}{4}, whose flux overlaps with the $A$~--~$X$ (0~--~0) band because of the width of the \ion{C}{4} line profile and the broad rotational state population of the CO. 
Alternatively, CO excitation by non-thermal electrons~\citep{beegle99}, analogous to collisionally excited H$_{2}$ emission (Bergin et al. (2004); Paper I), is an intriguing possibility.
 In disks orbiting stars with strong photospheric/chromospheric carbon emission, \ion{C}{1} $\lambda$1657~\AA\ may contribute to the pumping through a coincidence with the (0~--~2) CO band.   The analysis and modeling of this emission is ongoing, and a more complete analysis will be described in the larger CO survey. As noted in Paper I, the (0~--~1) $\lambda$1597~\AA\ feature is a major source of spectral confusion for studies of electron impact excited H$_{2}$ in CTTS observations at low spectral resolution (e.g., Ingleby et al. 2009). H$_{2}$ excited by electrons with energies, $E_{e}$,~$\gtrsim$~15 eV displays a characteristic gap in its emission spectrum at 1600~\AA. Thus, in order to use the 1600~\AA\ flux to determine the mass surface density of the inner molecular disk, the contribution from CO must be accounted for. This is complicated by the wealth of stronger photo-excited H$_{2}$ lines in the $\lambda$~$\leq$~1650~\AA\ bandpass. The 1712~--~1720~\AA\ spectrum, which is predominantly emission from the CO (0~--~3) and (14~--~12) bands, is free of Lyman and Werner band H$_{2}$ lines~\citep{abgrall93a} and can be used as a signpost for CO in the emission spectrum. 
We also detect spectral features coincident with the (1~--~1) $\lambda$1560~\AA\ and (1~--~3) $\lambda$1670~\AA\ bands. The latter is seen in the spectrum of HN Tau presented in Figure 3. The (1~--~1) band is coincident with stellar \ion{C}{1} $\lambda$1561~\AA\ and we propose that the (1~--~$v^{''}$) emission could be partially excited by \ion{C}{1} photons. 

\begin{figure}
\begin{center}
\epsfig{figure=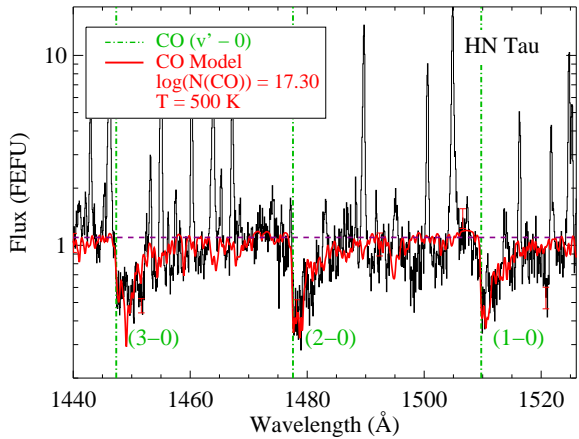,width=2.55in,angle=90}
\vspace{-0.2in}
\caption{The HN Tau spectrum in the 1440~--~1526~\AA\ bandpass is shown in black.
Fitting the CO transmission spectrum (fit displayed as the red line), we find $N$(CO)~=~2~$\pm$~1~$\times$~10$^{17}$ cm$^{-2}$, $T_{rot}$(CO)~=~500~$\pm$~200 K, suggesting that the absorbing gas is in the warm inner disk.  
The ($v^{'}$~--~0) absorption bands suggest a line of sight through much of the disk and thus a 
high disk inclination because the absorption bands are not observed in V4046 Sgr or RECX-11, whose CO emission suggests even higher column densities.  
The continuum level has been set to 1.2 FEFU over the 1440~--~1526~\AA\ band for this fitting.  
\label{cosovly}
 }
\end{center}
\end{figure}

\subsubsection{CO Fourth Positive Absorption Bands}

We detect CO absorption bands in the spectrum of HN Tau. 
The CO absorption bands arise from the ground vibrational state ($v^{'}$~--~0), and are seen superimposed on the continuum produced by the accretion shock. Figure 4 shows the absorption spectrum of HN Tau from 1440~--~1526~\AA\ where the strongest absorbers ($v^{'}$~=~1~--~3) are observed. 
The absorption profile can be explained by a viewing geometry where we are looking through an optically thick region of a high-inclination system, though not all high-inclination disk systems in the Taurus-Auriga region show warm CO absorption signatures (e.g., DF Tau).  
As discussed below, the wealth of H$_{2}$ and atomic emission lines makes a determination of the continuum level very challenging, so we assume a flat continuum level in this region of F$_{\lambda}$~=~1.2~$\times$~10$^{-15}$ erg cm$^{-2}$ s$^{-1}$ \AA$^{-1}$ ($\equiv$~1.2 FEFU). 
Assuming this continuum level in the 1440~--~1526~\AA\ bandpass, each absorption band has a width of 10~--~15~\AA, indicating a high degree of rotational excitation. 
For comparison, resolved absorption bands of cool CO ($T_{rot}$(CO)~$\approx$~4 K) characteristic of diffuse and translucent interstellar material have typical widths~$\lesssim$~0.5~\AA~\citep{burgh07}. The observed band widths imply that the CO rotational excitation temperature in HN Tau absorbing gas must be greater than a few~$\times$~10$^{2}$~K.  
Table 2 lists the absorption bands observed in HN Tau.

\subsection{Ly$\alpha$ and the Far-UV Continuum}

Figure 5 displays the Ly$\alpha$ profiles of the three systems. COS is a slitless spectrograph, and contamination from geocoronal \ion{H}{1} emission 
renders unusable the inner 1.2~--~3.0~\AA\ of the Ly$\alpha$ profile.
We can, however, measure the broad wings of the stellar+accretion Ly$\alpha$ profile, where the majority of the CO absorption takes place. In order to quantify the strength of the available Ly$\alpha$ pumping flux, we integrate each of the Ly$\alpha$ profiles from the CO (14~--~0) bandhead [$\lambda_{rest}$~=~1214.22~\AA] to 1220~\AA, masking out the airglow emission.  The extinction towards RECX-11 and V4046 Sgr is relatively small~\citep{luhman04,stempels04}, and from the observed Ly$\alpha$ profiles we estimate neutral hydrogen column densities of log($N$(H))~$<$~19.5 cm$^{-2}$, corresponding to $E(B~-~V)$~$<$~0.006 for typical interstellar gas-to-dust 
ratios~\citep{bohlin78}.  No dereddening was performed for V4046 Sgr or RECX-11.  
The extinction to HN Tau is somewhat uncertain, so we adopt the average 
of the extinction to the individual components ($A_{V}$~$\approx$~1.0; White \& Ghez 2001), and deredden the spectrum assuming the~\citet{ccm} curve for $R_{V}$~=~4.0~\citep{mathis90}. 
This value may underestimate the reddening associated with HN Tau; however, this value does not impact any of the CO parameters derived in \S3.3.  

\begin{deluxetable}{lccc}
\tabletypesize{\footnotesize}
\tablecaption{CO Fourth Positive bands identified in the spectra of CTTSs. \label{lya_lines}}
\tablewidth{0pt}
\tablehead{
\colhead{Band ID\tablenotemark{a}} & \colhead{$\lambda_{obs}$\tablenotemark{b}} & 
\colhead{Pumping Source} & \colhead{$\lambda_{pump}$} \\ 
& (\AA) & & (\AA) }
\startdata
Emission & & & \\
\tableline
(14~--~2) & 1280.5 & \ion{H}{1} Ly$\alpha$ & 1214~--~1220 \\
(14~--~3) & 1315.7 & \ion{H}{1} Ly$\alpha$ & 1214~--~1220 \\
(14~--~4) & 1352.4 & \ion{H}{1} Ly$\alpha$ & 1214~--~1220 \\
(14~--~5) & 1390.7 & \ion{H}{1} Ly$\alpha$ & 1214~--~1220 \\
(14~--~7) & 1472.6 & \ion{H}{1} Ly$\alpha$ & 1214~--~1220 \\
(14~--~8) & 1516.3 & \ion{H}{1} Ly$\alpha$ & 1214~--~1220 \\
(14~--~10) & 1610.1 & \ion{H}{1} Ly$\alpha$ & 1214~--~1220 \\
(14~--~12) & 1713.2 & \ion{H}{1} Ly$\alpha$ & 1214~--~1220 \\
(1~--~1) & 1560.2 & \ion{C}{1}, $e^{-}$ &  1561 \\
(1~--~3) & 1669.9 & \ion{C}{1}, $e^{-}$ &  1561 \\
(0~--~1) & 1597.3 & \ion{C}{4}, \ion{C}{1}, $e^{-}$ & 1544~--~1550, 1657 \\
(0~--~2) & 1653.2 & \ion{C}{4}, \ion{C}{1}, $e^{-}$ & 1544~--~1550, 1657 \\
(0~--~3) & 1712.4 &\ion{C}{4}, \ion{C}{1}, $e^{-}$ & 1544~--~1550, 1657 \\
\tableline
Absorption & & &  \\
\tableline
(0~--~0) & 1544.4 & $\cdots$ & $\cdots$ \\
(1~--~0) & 1509.8 & $\cdots$ & $\cdots$ \\
(2~--~0) & 1477.6 & $\cdots$ & $\cdots$ \\
(3~--~0) & 1447.4 & $\cdots$ & $\cdots$ \\
(4~--~0) & 1419.0 & $\cdots$ & $\cdots$ \\
(6~--~0) & 1367.6 & $\cdots$ & $\cdots$ \\
(7~--~0) & 1344.2 & $\cdots$ & $\cdots$ \\
(8~--~0) & 1322.1 & $\cdots$ & $\cdots$ 
\enddata
\tablenotetext{a}{Transitions are for the $A$$^{1}\Pi$~--~$X$$^{1}\Sigma^{+}$ CO band system. Band identifications for the emission
are labeled ($v^{'}$~--~$v^{''}$) and absorption lines are labeled ($v^{'}$~--~$v$). } 
 \tablenotetext{b}{CO wavelengths are taken from~\citet{kurucz93}.} 
\end{deluxetable}

We calculate the strength of the Ly$\alpha$ radiation field ($G_{Ly\alpha}$) at 1~AU from the central stars in units of the interstellar ultraviolet radiation field~\citep{habing68}. While this treatment does not make any correction for the ISM or the geometry of the inner disk, we find that the flux of the Ly$\alpha$ pumped CO emission is proportional to the strength of the Ly$\alpha$ profile as observed at Earth (columns 6 and 7 in Table 3).  We note that the same is not necessarily true for H$_{2}$ (e.g., we see numerous fluorescent H$_{2}$ emission lines in the spectrum of HN Tau without directly observing strong Ly$\alpha$ emission).  
We have not attempted to reconstruct the stellar Ly$\alpha$ profile here. We can compare the observed Ly$\alpha$ emission strength with that of the continuum, $G_{cont}$.  Following the method described in Paper I, we have measured the  continuum spectra of the three systems. These data are shown in Figure 6, where the flux units have been scaled to the local radiation field strengths at 1 AU from their host stars. 
The integrated 912~--~2000~\AA\ fluxes are presented in Table 3. We find that the observed $G_{Ly\alpha}$ and $G_{cont}$ are comparable for V4046 Sgr and RECX-11, while the strength of the UV continuum is more than two orders of magnitude greater than the observed (unreconstructed) Ly$\alpha$ flux in HN Tau. 
The large difference between the Ly$\alpha$ and accretion continuum observed in HN Tau is most likely attributable to strong \ion{H}{1} attenuation on the sightline (disk + interstellar) to HN Tau.  The Ly$\alpha$ photons are scattered out of our line-of-sight while passing through the disk, whereas the continuum emission is only attenuated by dust grains.  The grain opacity is several orders of magnitude less than
the line-center atomic hydrogen opacity~\citep{fogel11}.  Additionally, 
grain growth acts to reduce the UV dust opacity~\citep{vasyunin10}, making the continuum emission more readily observable through the disk.  

\begin{figure}
\begin{center}
\epsfig{figure=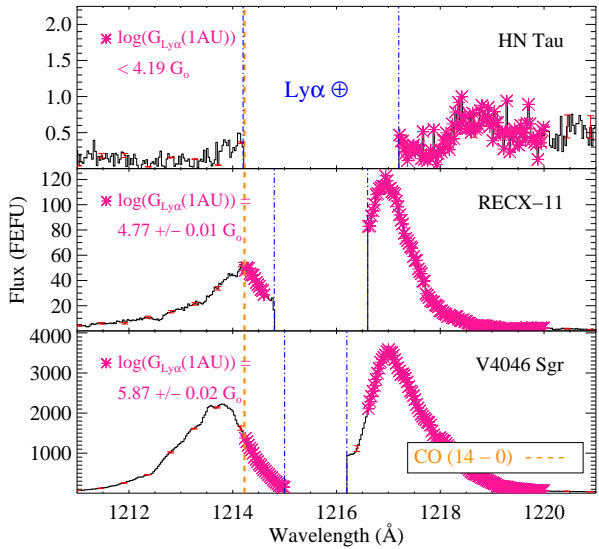,width=2.55in,angle=90}
\vspace{+0.2in}
\caption{The strength of the $observed$ Ly$\alpha$ emission across the CO (14~--~0) band is proportional to the strength of the resultant fluorescence (compare e.g., with Figure 1).  We have calculated the strength of the radiation field available to be absorbed at 1 AU (in units of the Habing flux, $G_{o}$~=~1.6~$\times$~10$^{-3}$ erg cm$^{-2}$ s$^{-1}$), at wavelengths unaffected by geocoronal emission ({\it magenta stars}).  The integrated flux level is shown at the upper left of each panel and in Table 3.  The inner regions of the HN Tau profile are 
still contaminated by the wings of the geocoronal \ion{H}{1}, thus only an upper limit to the Ly$\alpha$ emission can be determined.  The position of the (14~--~0) bandhead is marked with the dashed orange line.  
\label{cosovly} 
 }
\end{center}
\end{figure}

\subsection{Analysis} 

\subsubsection{Ly$\alpha$ Pumping of CO}

We constructed a simple model to estimate the properties  
of the photo-excited CO population (the CO column density on the star-disk sightline, $N$(CO), and the CO rotational temperature, $T_{rot}$(CO)) in the inner disks of V4046 Sgr and RECX-11. We used the Dunham coefficients from~\citet{george94} to calculate the ground electronic state CO term values.  The ground state energy is used to create a population distribution ($P(v=0,J)$) that is characterized by $T_{rot}$(CO). 
We then compute the absorption cross-sections for the $A$~--~$X$ (14~--~0) transitions using the Einstein $A$-values and wavelengths from~\citet{kurucz93}. The CO optical depth is $\tau_{\lambda}$~=~$N$(CO)$P(0,J)$$\sigma_{\lambda}$. The total absorbed flux is then 
\begin{equation}
I_{\lambda}~=~ I^{o}_{\lambda}(1 - e^{-\tau_{\lambda}}) 
\end{equation}
where $ I^{o}_{\lambda}$ is the incident Ly$\alpha$ radiation field. Our model does not make an assumption about the geometry of the absorbing gas.  Because the transition probabilities of the (14~--~0) absorption lines are intrinsically small ($\sim$~10$^{4}$ s$^{-1}$), optical depth effects do not dominate the uncertainty on the total absorbed/emitted flux.  A spatial distribution where the CO is concentrated in one dense parcel at the inner edge of the molecular gas disk cannot be distinguished from a more tenuous distribution of CO across a larger disk surface in our approach.  

The emitted flux is determined by the branching ratios to the various rovibrational levels of the $X$$^{1}\Sigma^{+}$ ground electronic state. The branching ratios for a given transition ($A$$^{1}\Pi$,$v^{'}$,$J^{'}$) $\rightarrow$ ($X$$^{1}\Sigma^{+}$,$v^{''}$,$J^{''}$) are subject to the appropriate dipole selection rules and the Franck-Condon factors. The crucial input for this simplified model of the observed CO emission is $ I^{o}_{\lambda}$, the stellar+shock Ly$\alpha$ profile as seen by the CO in the disk. The shape of this profile will depend strongly on the intervening neutral hydrogen in the protostellar outflow and disk atmosphere. The intervening \ion{H}{1} absorbers will not only determine the absolute level of the incident radiation field, but also the wavelengths of accessible pumping photons. For simplicity, we assume that the observed Ly$\alpha$ profile, $not$ including the region near the line center where interstellar and geocoronal \ion{H}{1} complicate the profile, is representative of the Ly$\alpha$ seen by the CO.  
In order to create a continuous Ly$\alpha$ pumping profile, 
we fit a parabola across the spectral region where geocoronal emission has been removed. 

\begin{figure}
\begin{center}
\epsfig{figure=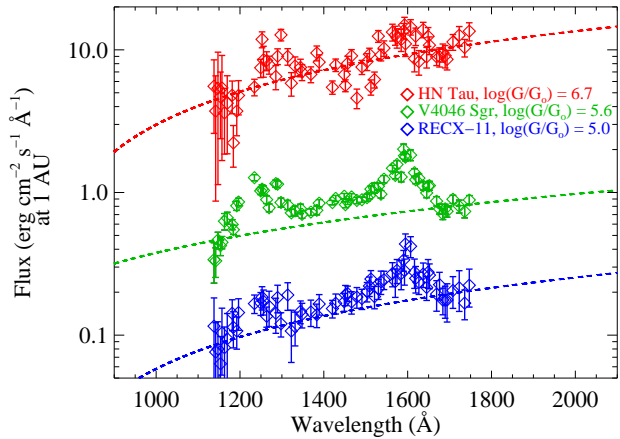,width=2.55in,angle=90}
\vspace{-0.2in}
\caption{The far-UV continuum in our three targets.  The continuum is measured in 0.75~\AA\ bins between stellar and disk emission lines (Paper I).  The binned continuum measurements are shown as diamonds.  The HN Tau data has been corrected for interstellar reddening, adopting the average extinction of the HN Tau system~\citep{white01}, while RECX-11 and V4046 Sgr are consistent with $E(B~-~V)$~$<$~0.006 (\S 3.2) and no reddening correction is performed. 
The dashed lines are fits to the spectra in regions that are free of contamination from photo-excited H$_{2}$ emission ($\lambda$~$<$~1180 and $\lambda$~$>$~1660~\AA).  The smooth emission above the dashed lines is attributable to collisionally excited H$_{2}$ (Bergin et al. 2004; Paper I).  
The continuum fits are extrapolated over the 912~--~2000~\AA\ bandpass, and the integrated fluxes are listed in the legend and  Table~3.
\label{cosovly} 
 }
\end{center}
\end{figure}

We illustrate the modeling process for V4046 Sgr in Figure 2. 
In the top panel, we show the wavelength distribution of rotational states in the (14~--~0) band. 
In the middle panel, we show the stellar Ly$\alpha$ profile, with the flux absorbed by CO in green. We note that the CO absorption is not seen against the Ly$\alpha$ profile, unlike the H$_{2}$ absorption profiles discussed by~\citet{yang11}. The lack of observable CO absorption on the Ly$\alpha$ profile is most likely due to a combination of the low-inclination of the system, the relatively small transition probabilities from the ground vibrational level to $v^{'}$~=~14, and the moderate resolving power of COS.   
The absorbed flux is redistributed among the $v^{''}$ levels during the fluorescent cascade, and we show the (14~--~3) emission band in the lower panel of Figure 2. The model emission is shown in green. States affected by uncertainties in the Ly$\alpha$ profile are marked in gray.  Figure 2 shows that the model underpredicts the observed CO emission for the rotational states that are pumped by the Ly$\alpha$ line center. This indicates that the illumination by the Ly$\alpha$ line center is significant in the inner disk, consistent with the observation of strong H$_{2}$ emission pumped through the $B$~--~$X$ (1~--~2) R(6) 1215.73~\AA\ and (1~--~2) P(5) 1216.07~\AA\ absorbing transitions.  In the larger CTTS CO survey in preparation, we will present a more sophisticated CO fluorescence model, including a sefl-consistent reconstruction of the Ly$\alpha$ profile.

\begin{deluxetable*}{cccccccccc}
\tabletypesize{\scriptsize}
\setlength{\tabcolsep}{0.04in} 
\tablecaption{Disk CO Parameters. \label{cos_params}}
\tablewidth{0pt}
\tablehead{
\colhead{Object} & \colhead{Spectral Type} & \colhead{Region} & \colhead{Inclination} & \colhead{Ref.} & \colhead{log(F(CO))\tablenotemark{a}}
& \colhead{log($G_{Ly\alpha}$/$G_{o}$)\tablenotemark{b}} & \colhead{log($G_{cont}$/$G_{o}$)} & 
\colhead{log$N$(CO)} & \colhead{$T_{rot}$(CO)} \\
\colhead{ } & \colhead{ } & \colhead{} & \colhead{ } & \colhead{} & \colhead{(erg cm$^{-2}$ s$^{-1}$)} & \colhead{($a$~=~1 AU)} & \colhead{($a$~=~1 AU)} & \colhead{(cm$^{-2}$)} & \colhead{(K)} 
}
\startdata 
RECX-11 & K5.5 & $\eta$ Cha & 70\arcdeg & {\it 1,2,3} & -13.70 & 4.8 & 5.0 & 18.9$^{+0.3}_{-0.4}$ & $>$~200K \\
HN Tau & K5 + M4 & Taurus & $>$~40\arcdeg & {\it 4} & $<$ -14.07 & $<$ 4.2 & 6.7 & 17.3$^{+0.2}_{-0.3}$\tablenotemark{c} & 500~$\pm$~200\tablenotemark{c} \\
V4046 Sgr & K5 + K5 & isolated & 35\arcdeg & {\it 5,6,7,8} & -12.44 & 5.9 & 5.6 & 
18.75$^{+0.15}_{-0.25}$ & 290~$\pm$~90 
\enddata
\tablenotetext{a}{Total integrated CO flux pumped by stellar+accretion shock Ly$\alpha$ photons.}
\tablenotetext{b}{$G_{o}$ is the average interstellar radiation field evaluated over the 912~--~2000~\AA\ bandpass ($G_{o}$~=~1.6~$\times$~10$^{-3}$ erg cm$^{-2}$ s$^{-1}$; Habing 1968).\nocite{habing68} } 
\tablenotetext{c}{Determined from the CO absorption on the line of the sight through the disk. \\
References: (1)~\citet{mamajek99}, (2)~\citet{luhman04}, (3)~\citet{lawson04}, 
(4)~\citet{white01}, (5)~\citet{quast00}, (6)~\citet{stempels04}, (7)~\citet{kastner08}, (8)~\citet{rodriguez10} } 
\end{deluxetable*}

$N$(CO) in the ground electronic state is determined by the amount of absorption required to produce the observed fluorescent emission in our approach.  For a completely optically thin parcel of gas, $N$(CO) is degenerate with the covering fraction of CO in the disk.   
The rotational excitation temperature controls the rotational-state population of the absorbing molecules, which we constrain by observing the distribution of emission lines in highly-excited rotational levels ($J^{''}$~$>$~10). We searched a grid of column density and temperatures for the best fit to the COS observations. Exploring regions of parameter-space that are inconsistent with the observations allows us to place error bars on $N$(CO) and $T_{rot}$(CO).  Figure 7 shows synthetic CO spectra for the 1-$\sigma$ error bars of $T_{rot}$(CO) to illustrate the model dependence on the temperature.  
Values for $N$(CO) that are outside of the error bars predict too much or too little total CO flux, while values for $T_{rot}$(CO) that are outside of the error bars either do not predict the high-$J$ CO lines seen in the data or predict many more lines than are observed.  
We note that the highest rotational levels may not be thermalized due to their larger critical densities ($n$(H)$_{crit}$~$\gtrsim$~10$^{5}$ cm$^{-3}$ for $J$~$>$~10).  This effect would cause our $T_{rot}$(CO) to systematically underestimate the kinetic temperature of the CO-bearing gas.   
Due to the much stronger relative contribution of geocoronal Ly$\alpha$ and the lower S/N of the RECX-11 spectrum, we were only able to use the red side of the pumping profile and thus higher rotational levels to constrain the CO parameters in this system.  The uncertainties on $N$(CO) are correspondingly larger for RECX-11.  
With this simple approach, we can only set a lower limit on the rotational excitation temperature because the red-wing of the RECX-11 Ly$\alpha$ profile does not excite 
CO with $J$~$\gtrsim$~20, so higher rotational temperatures cannot be excluded. 

Following this modeling procedure, we determine that the Ly$\alpha$-pumped CO emission observed in V4046 Sgr and RECX-11 has column densities of $N$(CO)~=~5.6~$\pm$~2.3~$\times$~10$^{18}$ cm$^{-2}$ and 8~$\pm$~4~$\times$~10$^{18}$ cm$^{-2}$, respectively. The rotational excitation temperatures for V4046 Sgr and RECX-11 are $T_{rot}$(CO)~=~290~$\pm$~90 K and $>$~200 K.  
A velocity broadening of 20 km s$^{-1}$ was applied to the emission spectrum to match the width of the high-$J$ lines observed in the V4046 Sgr data.  Due to its large molecular mass, the thermal width of CO lines is relatively small ($v_{therm}$~$<$~1 km s$^{-1}$ for $T_{rot}$(CO)~$<$~1700 K), so the velocity broadening is almost certainly dominated by a combination of turbulence and Keplerian rotation of the CO-bearing gas.  It should be noted that this is smaller than the 52 km s$^{-1}$ line width observed in the sample of 13 photo-excited H$_{2}$ lines from V4046 Sgr presented in Paper I, however a direct comparison of these values should be made with caution due to blending between the R, P, and Q branches as well as uncertainties in the CO wavelengths for the high-$J$ (14~--~$v^{''}$) lines.

\subsubsection{CO Absorption}

We can also use the CO ($v^{'}$~--~0) absorption spectrum to constrain the column density and temperature of the CO in high-inclination disks. 
We have created synthetic optical depth spectra for the attenuation of UV photons through a screen of CO molecules to compare with the HN Tau spectra. These optical depths assume a turbulent velocity (0.1~km s$^{-1}$) and are characterized by a column density and rotational temperature. The model CO absorption spectrum is compared to the observed CO absorption bands in Figure 4. There are two main points of uncertainty with this treatment of the absorption lines: 1) the continuum placement and 2) the S/N ratio of the data. There are many emission lines from photo-excited H$_{2}$ and hot gas in the far-UV spectra of CTTS that make a determination of the continuum level challenging. Continuum placement is critical for profile-fitting of molecular absorption spectra~\citep{sonnentrucker07}. In the case of HN Tau, uncertainties on the continuum level complicate measurements of high-$J$ absorption (and thus $T_{rot}$). 
The flux in the HN Tau spectra in the region of CO absorption is F$_{\lambda}$~$\lesssim$~1 FEFU, so structure in the line profiles is lost in the noise. 
Despite these challenges, we can use the ($v^{'}$~--~0) absorption bands to constrain the column density and temperature to $N$(CO)~=~2~$\pm$~1~$\times$~10$^{17}$ cm$^{-2}$ and $T_{rot}$(CO)~=~500~$\pm$~200~K. 

\begin{figure}
\begin{center}
\epsfig{figure=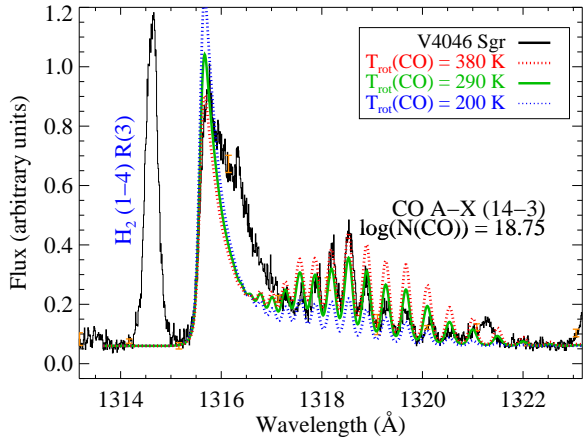,width=2.55in,angle=90}
\vspace{-0.2in}
\caption{CO models for the $T_{rot}$(CO) consistent with the COS observations of V4046 Sgr ({\it shown in black}, see Figure 2).  There is a slight $N$--$T_{rot}$ degeneracy in the model fits, though in general, $N$(CO) controls the output flux and $T_{rot}$(CO) controls the population of the high-$J$ states seen in the CO emission profiles.
The best fit models find $N$(CO)~=~5.6~$\pm$~2.3~$\times$~10$^{18}$ cm$^{-2}$ and $T_{rot}$(CO)~=~290~$\pm$~90~K.   The model underpredicts the observed flux between 1316~--~1317~\AA\ due to uncertainties in the exciting Ly$\alpha$   
emission profile (\S3.3.1). 
\label{cosovly} 
 }
\end{center}
\end{figure}


\section{Discussion of Far-UV CO and H$_{2}$ Emission from the Inner Disk}

\subsection{The Molecular Excitation and Spatial Distributions}

We have used the combination of sensitivity and spectral resolution afforded by $HST$-COS to isolate and characterize CO emission in the far-UV spectra of protoplanetary disks for the first time. Our values for the rotational temperature are in the range of a few hundred degrees K, and if we assume that the CO population is in thermal equilibrium with the stellar radiation, this places the CO at an orbital distance of $\sim$~0.2~--~2 AU in our targets. Additionally, CO rotational temperatures in this range suggests that the photo-excited CO we observe may not be related to the CO  traced through the $\lambda$~$\sim$~4.7~$\mu$m fundamental bands, although this correlation will require more study.  
In general, we find that the fluorescent CO emission detected here arises in gas that is somewhat less energetic than traced by $M$-band CO emission, which is characterized by temperatures of $\sim$~1000 K~\citep{najita03}.   
It should be noted however that our results are roughly consistent with the rotational temperatures measured from a sample of VLT-CRIRES spectra of protoplanetary disks~\citep{bast11}. These authors find a decoupling of the rotational and vibrational populations in a small subsample of CTTSs with large accretion rates, suggesting that the ground state vibrational populations are influenced by non-thermal UV photo-excitation (see also Brittain et al. 2007).\nocite{brittain07}
 Interestingly, an analogous process is may be occurring in the H$_{2}$ population of the inner disk, namely that excitation by non-thermal processes such as photo-excitation~\citep{ardila02} and excitation by photoelectrons~(Bergin et al. 2004; Paper I) could lead to an effective vibrational temperature that is significantly higher than the rotational or kinetic temperature of the molecules.

At the temperatures and densities characteristic of the inner gas disks, the small critical densities of the pure rotational transitions of H$_{2}$ ($n_{crit}$~$<$~2000 cm$^{-3}$ for $J^{''}$~$\leq$~3 and $T_{rot}$(H$_{2}$)~$\gtrsim$~1000 K; Mandy \& Martin 1993) 
should should make them excellent tracers of the kinetic temperature.~\nocite{mandy93} However, due to their very small transition probabilities, H$_{2}$ rotational lines have proven very difficult to observe in circumstellar disk environments~\citep{pascucci06,lahuis07,bitner08,carmona08}. Larger aperture and source sizes allowed for robust detections of the H$_{2}$ rotational spectrum in photodissociation regions (PDRs) with the $Spitzer$-IRS (e.g., Fleming et al. 2010). PDRs and the inner regions of protoplanetary disks display qualitative similarities: a strong UV radiation field photodissociating and partially ionizing a molecular medium~\citep{gorti02,gorti08}. Far-UV observations of photo-excited H$_{2}$ in PDRs show a decoupling of the vibrational and rotational temperatures, with vibrational temperatures of order 2000~--~3000 K~\citep{france05a}. Mid-IR observations of these same regions with $ISO$, $Spitzer$, and ground-based spectrographs reveal that the rotational temperatures of these regions are typically 300~--~900~K~\citep{habart04,allers05,france07a,fleming10}. This is the same qualitative behavior observed in CO observations of CTTS disks, and we argue that the UV-pumping process can influence both the CO and H$_{2}$ level populations in these systems. 

The similarity of the physical conditions in the H$_{2}$ and CO emitting regions, particularly the fact that both species show photoexcitation from the Ly$\alpha$ line core (\S3.3.1), naively implies that they are co-spatial. 
Deep far-UV spectroscopy allows us to observe both molecular species from a common physical origin in a single observation.  The argument we present here is somewhat qualitative, the physical state and composition of the molecular inner disk is a complicated balance of radiative excitation by continuum and line photons, collisional population of the rovibrational levels of the ground electronic state, and non-thermal processes such as collisions with photoelectrons.  A rigorous treatment of the distribution of photo- and collisionally-excited molecular populations is beyond the scope of this work.  More complex theoretical models that predict both the UV and sub-mm/mm wavelength line fluxes are needed for future studies of the warm molecular phase of protoplanetary disks.

The fact that the Ly$\alpha$-pumping of CO proceeds at all is a strong argument for a disk surface origin of the CO emission.  In the absence of a local source of Ly$\alpha$ photons within the disk, our observations strongly suggest that both the CO and H$_{2}$ are shielded by very little neutral hydrogen. 
If one assumes a standard dark cloud $N$(CO)/$N$(H$_{2}$) ($\equiv$~CO/H$_{2}$~$\approx$~10$^{-4}$) conversion factor and a molecular fraction of 0.67, CO column densities of order 10$^{18}$ cm$^{-2}$ would be located in a medium with $N$(H)~$\sim$~10$^{22}$ cm$^{-2}$. This is clearly not feasible  because the average neutral hydrogen optical depth at the wavelengths required to excite the observed CO fluorescence ($<$$\tau_{HI}$(1210~--~1220~\AA)$>$) would be $>$~10$^{7}$ at $N$(H)~$\sim$~10$^{22}$ cm$^{-2}$. \ion{H}{1} column densities larger than $\sim$~10$^{20}$ cm$^{-2}$ are incompatible with the observed CO emission.  This implies that CO/H~$>$~10$^{-2}$ in the CO emitting region and argues strongly for an origin at the disk surface.

\subsection{The CO/H$_{2}$ Ratio in the Inner Disk}

Having demonstrated that the CO and H$_{2}$ may be sampling the same distribution of inner disk gas, we can evaluate the CO/H$_{2}$ ratio, $X_{CO}$~\citep{liszt10}, in the inner disk.  The photo-excited $N$(H$_{2}$) in the inner disk of CTTSs has been measured in the range 10$^{18-19}$ cm$^{-2}$ (Herczeg et al. 2004; and extrapolating the high excitation H$_{2}$ column densities from Ardila et al. 2002). This is essentially identical to the column density of H$_{2}$ excited by electron impact in CTTS disks (Ingleby et al. 2009; Paper I), although it is not yet clear whether or not collisionally excited gas traces the photo-excited population. 
Thus, we suggest that the CO/H$_{2}$ ratio in the disk surface at terrestrial planet forming radii is of order unity ($N$(CO)/$N$(H$_{2}$)~$\sim$~1), far from the dense cloud value of 10$^{-4}$ that is typically assumed. 

CO/H$_{2}$~$\sim$~1 suggests considerable evolution from the dense cloud from which the system formed, and may represent a transitional phase between the interstellar and cometary environments (CO/H$_{2}$~$\sim$~30; Lupu et al. 2007)~\nocite{lupu07} where water ice regulates the gas phase abundances of both molecules. We suggest two possible processes that could drive the CO/H$_{2}$ ratio towards the observed levels. Photoevaporation and/or collisions of large grains/planetesimals have been suggested as sources for maintaining a gas-phase carbon abundance in debris disk systems~\citep{vidal94,lecavelier01,chen03}. We suggest that this second-generation CO could be produced in collisions in the dust disk region, and transported to the surface of the inner disk by mass-flow induced by photoevaporation~\citep{gorti08}. 

H$_{2}$ could be selectively dissociated by the radiation field present in the inner disk.  H$_{2}$ can be photodissociated by a two-step process where a photon is absorbed into the Lyman band system followed by emission into the vibrational continuum, when $v^{''}$~$>$~14~\citep{stecher67}. For molecules exposed to the far-UV continuum of an O or B star, the dissociation probability ($p_{diss}$~=~$A_{cont}$/$A_{TOT}$, the ratio of the transition probability to the vibrational continuum compared with the total transitional probability from a given upper state) is 0.1~--~0.15.   Warm H$_{2}$ will have absorption out of the high-$J$ lines of the Lyman (0~--~0) band at wavelengths as long as 1140~--~1150~\AA, and multiple pumping effects can introduce additional absorption opacity at longer wavelengths. 
The predissociating transitions of CO all require photons of $\lambda$~$\lesssim$~1076~\AA\  (starting with the $E$~--~$X$ bands; van Dishoeck \& Black 1988).\nocite{vand88}
Predissociation of CO through the $A$~--~$X$ bands is negligible (see e.g., the $A$~--~$X$ system parameters measured by Le Floch et al. 1987).\nocite{lefloch87} 
We have shown that the accretion-powered far-UV continuum is decreasing from 1150~--~1076~\AA\  (Figure 6), indicating that H$_{2}$ molecules in circumstellar disks may be selectively dissociated by continuum photons more readily than in interstellar clouds. 
However, the dominant mechanism for H$_{2}$ dissociation in the inner disk is most likely the huge flux of Ly$\alpha$ photons produced by the star+accretion shock. The Lyman band H$_{2}$ transitions in the range 1214~--~1217~\AA\ have very low dissociation probabilities. However, there are several Werner band lines in that region with both large radiative rates into the upper level and large rates to the continuum ($p_{diss}$~$\sim$~0.07~--~0.17). Furthermore, there are several strong transitions ($A_{TOT}$~$>$~10$^{8}$ s$^{-1}$) to the $B^{'}$ electronic levels with $p_{diss}$~$>$~0.5 coincident with Ly$\alpha$. A complete characterization of the abundances of CO and H$_{2}$ in the inner dust disk will require additional modeling, but mechanisms to produce high CO/H$_{2}$ ratios are available. 

Alternatively, there may be a geometrically thin population of H$_{2}$ characterized by kinetic temperatures of order 2500 K~\citep{herczeg04}. If this population dominates the observed H$_{2}$ fluorescence, then the CO is tracing a cooler bulk distribution where the $N$(H$_{2}$) may be higher.  In this case, a direct comparison between the two molecules breaks down.
Finally, the CO/H$_{2}$ ratio observed in the inner disk does not necessarily have any bearing on the CO/H$_{2}$ ratio in the more quiescent outer disk probed by mm-wave CO measurements. However, our results suggest that the traditional assumption of a dense cloud $X_{CO}$-conversion factor for determining the total molecular mass of a protoplanetary disk may not be correct, and more work should be focused on determining an environmentally-dependent $X_{CO}$-conversion factor.

\section{Summary}

We have presented the first far-ultraviolet detections of CO emission and absorption from the planet-forming region of CTTS disks.    We detect for the first time Ly$\alpha$-pumped CO emission bands in V4046 Sgr and RECX-11.  
The far-UV emitting CO is characterized by $N$(CO)~$\sim$~10$^{18-19}$ cm$^{-2}$, with rotational excitation temperatures of $\gtrsim$~300~K. 
We clearly detect warm CO in absorption against the accretion continuum in HN Tau with $N$(CO)$\sim$~2~$\times$~10$^{17}$ cm$^{-2}$, $T_{rot}$(CO)~$\sim$~500~K.  
These data provide direct evidence for the influence of UV photons on CO-rich inner disks.
We compare these results with observations of photo- and collisionally-excited H$_{2}$ to  determine that the CO and H$_{2}$ populations observed with far-UV spectroscopy may be co-spatial. This suggests that the CO/H$_{2}$ ratio in the inner regions of protoplanetary disks is of order unity, possibly due to a combination of replenishment of gas-phase CO and selective photodissociation of H$_{2}$.  We are in the process of assembling a larger survey of photo-excited CO emission in protoplanetary disk targets.  This work, which will present more sophisticated modeling of the Ly$\alpha$ radiation field and CO fluorescence, will be used to compare the physical characteristics of the UV-emitting CO with system parameters such as age, mass accretion rate, and disk mass.  \\

{\bf Author Affiliations:} \\
\footnotesize
1~--~Center for Astrophysics and Space Astronomy, University of Colorado, 389 UCB, 
Boulder, CO 80309; kevin.france@colorado.edu \\
2~--~Max-Planck-Institut f\"{u}r extraterrestriche Physik, Postfach 1312, 85741 Garching, Germany \\
3~--~School of Physics, Trinity College, Dublin 2, Ireland \\
4~--~JILA, University of Colorado and NIST, 440 UCB, Boulder, CO 80309 \\
5~--~LUTH and UMR 8102 du CNRS, Observatoire de Paris, Section de Meudon, Place J. Janssen, 92195 Meudon, France \\
6~--~NASA Herschel Science Center, California Institute of Technology, 1200 E. California Blvd, Pasadena, CA 91125, USA \\
7~--~Department of Astronomy, University of Michigan, 830 Dennison Building, 500 Church Street, Ann Arbor, MI 48109, USA \\
8~--~NSF Astronomy and Astrophysics Postdoctoral Fellow.  Harvard-Smithsonian Center for Astrophysics, 60 Garden Street, MS-78, Cambridge, MA 02138, USA \\
9~--~California Institute of Technology, Department of Astrophysics, MC 249-17, Pasadena, CA 91125, USA \\
10~--~ESO, Karl-Schwarzschild-Strasse 2, 85748 Garching bei M\"{u}nchen, Germany \\
11~--~Department of Physics \& Astronomy, Rice University, Houston, TX 77005, USA \\
12~--~Space Telescope Science Institute, 3700 San Martin Dr, Baltimore MD 21218, USA \\
13~--~Stony Brook University, Stony Brook NY 11794-3800, USA \\
\normalsize

\acknowledgments
K. F. thanks Roxana Lupu for enjoyable discussions regarding the far-UV spectrum of CO. 
This work was support by NASA grants NNX08AC146 and NAS5-98043 to the University of Colorado at Boulder and STScI grants to program GO-11616.

\bibliography{ms_emapj_co}


\end{document}